# Photon-Induced Magnetization Reversal in Single-Molecule Magnets

M. Bal[1], Jonathan R. Friedman[1,*], Y. Suzuki[1,2], K. Mertes[1,2,†], E. M. Rumberger[3], D. N. Hendrickson[3], Y. Myasoedov[4], H. Shtrikman[4], N. Avraham[2,4] & E. Zeldov[4]

[1]*Department of Physics, Amherst College, Amherst, Massachusetts 01002-5000, USA*

[2]*Physics Department, City College of the City University of New York, New York, New York 10031, USA*

[3]*Department of Chemistry and Biochemistry, University of California at San Diego, La Jolla, California 92093, USA*

[4]*Department of Condensed Matter Physics, The Weizmann Institute of Science, Rehovot 76100, Israel*

**Abstract**

We use millimeter wave radiation to manipulate the populations of the energy levels of a single crystal molecular magnet Fe8. When a continuous wave radiation is in resonance with the transitions from the ground state to the first excited state, the equilibrium magnetization exhibits a dip. The position of this dip varies linearly with the radiation frequency. Our results provide a lower bound of 0.17 ns for transverse relaxation time and suggest the possibility that single-molecule magnets might be utilized for quantum computation.

Single-molecule magnets lie at the frontier between the quantum and classical worlds. Like classical magnets used for magnetic storage, they are bistable, exhibiting hysteresis at low temperatures [1,2]. However, they also exhibit striking quantum mechanical properties such as tunneling between "up" and "down" orientations [3-5] as well as a geometric-phase effect created by the interference between tunneling paths [6]. Recent experiments show that the rate for magnetization reversal can be augmented using microwave radiation [7,8]. Here we demonstrate a novel method of reversing the magnetization of a system of single-molecule magnets in which radiation at a resonant frequency induces a partial population inversion. These results open up the possibility that single-molecule magnets can be used for magnetic storage and as qubits [9], the processing elements in quantum computers.

The single-molecule magnet $Fe_8O_2(OH)_{12}(tacn)_6$ (heretofore called $Fe_8$) is composed of eight magnetic Fe(III) ions strongly coupled together to form a single spin-10 system. The molecules crystallize into a triclinic lattice and have a large biaxial magnetocrystalline anisotropy. Reversal of the magnetization from one easy-axis direction to another is impeded by a ~25 K barrier [5], suggesting a double-well model of the system's energy, as illustrated in Fig. 1, where the left (right) well corresponds to the spin pointing along (antiparallel to) the easy axis. The potential contains a series of energy levels that roughly correspond to different orientations of the magnetic moment. A magnetic field applied along the easy axis tilts the potential, favoring, e.g., the spin-up orientation. At certain values of magnetic field, levels in opposite wells will align, permitting resonant tunneling between up and down states, a phenomenon first discovered in the material Mn$_{12}$ acetate [3].

The system can be described by an effective spin Hamiltonian

$$\mathcal{H} = -DS_z^2 + E\left(S_x^2 - S_y^2\right) + C\left(S_+^4 + S_-^4\right) - g\mu_B \vec{S} \cdot \vec{H}, \tag{1}$$

where the anisotropy constants $D$, $E$, and $C$ are 0.292 K, 0.046 K, and -2.9 x 10$^{-5}$ K, respectively, and $g = 2$ [6,10,11]. The first (and largest) term causes the spin to prefer to lie along or opposite the z axis, resulting in the double-well potential of Fig. 1, and making the energy levels in Fig. 1 approximately the eigenstates of $S_z$. The second and third terms break the rotational symmetry of the Hamiltonian and result in tunneling between the otherwise unperturbed states. When the magnetic field $H$ is in the x-z plane, the Zeeman term can be rewritten as

$$g\mu_B \vec{S} \cdot \vec{H} = g\mu_B H\left(S_z \cos\theta + S_x \sin\theta\right), \tag{2}$$

where $\theta$ is the angle between the spin $S$ and the external magnetic field $H$. (In the experiments the field is actually ~17° out of the x-z plane [12], but this fact has only minor effects on the results presented.)

We mounted a single crystal of $Fe_8$ on a Hall-bar detector with its easy axis tilted ~33(1)º in the a-b plane from the field direction [12]. A component of the field perpendicular to the easy (z) axis enhances tunneling. The sample was irradiated with monochromatic microwaves. We measured the steady-state magnetization of the sample as a function of magnetic field both with and without the presence of radiation. (Since the radiation also heats the sample somewhat, we took great care to measure the magnetization curve in the absence of radiation at the same temperature as when the sample is irradiated. We could determine the temperature of the sample to within 1 mK from the magnetic relaxation rate measured with AC susceptibility when the DC magnetic field is far from the resonance peaks shown in Fig. 2. The relaxation rate, the frequency at which the in-phase component of the susceptibility inflects, is exponentially sensitive to changes in temperature). After subtracting the two curves we obtain $\Delta M$, the radiation-induced change in magnetization, as a function of magnetic field.

Fig. 2 shows $\Delta M$ as a function of field for several frequencies of radiation. Each curve shows that the radiation induces a change in the sample's magnetization at certain values of magnetic field. At these fields, the frequency of the radiation matches the energy difference between the lowest two levels in, e.g., the right well in Fig. 1, resulting in the absorption of a photon and subsequent thermal or tunneling relaxation (or a combination of both) into the left well. We find that the magnitude of the magnetization change is largest when the first excited state in the right well is near a tunneling resonance with a level in the opposite well, as in Fig. 2a. This indicates that the photon-induced reversal process can be enhanced by tunneling, consistent with the relaxation results of Sorace et al [8]. Our results show magnetization reversal even when levels in opposite wells are far from resonance, when tunneling is effectively nil. In this case, the radiation produces a nonthermal population in the first excited state in the right well. Thermal phonons then produce transitions between levels until a quasi-equilibrium is established, resulting in an increased population in the left well.

Some of our data show an asymmetry between peaks in negative field and those at positive field, as in Fig. 2a. We attribute this to elliptical polarization of the radiation produced by mode mixing in our waveguide. The asymmetry only affects the height of the peaks, but not their position.

The Zeeman term in the Hamiltonian implies that the energy difference between levels should vary linearly with external field. In Fig. 3 we plot the field at which magnetization reversal occurs as a function of microwave frequency (using the data shown in Fig. 2 and similar curves at other frequencies). We indeed obtain a linear dependence. The straight solid line in the figure results from numerically calculating the energy difference between the lowest two levels using the accepted Hamiltonian and anisotropy constants and setting $\theta = 34°$, a value that gives a good fit to the data and is also in agreement with the directly measured angle between the sample's easy axis and the field direction.

We numerically modeled our results by constructing a master rate equation that includes the spin-phonon transitions as well as photon-induced transitions. To do this, we first diagonalized the Hamiltonian, Eq. (1), and used the energy eigenstates as the basis for our master equation. Using this basis simplifies calculations, although it treats resonant tunneling as coherent. Since neither the experiment nor the calculations are done when the system is tuned precisely to resonance, the unphysical assumption of coherence does not present a problem.

The master rate equation we solved numerically is

$$\frac{dP_i}{dt} = -\sum_{\substack{j=1 \\ i \neq j}}^{21} \left( \gamma_{i,j}^{1+} + \gamma_{i,j}^{1-} + \gamma_{i,j}^{2+} + \gamma_{i,j}^{2-} + w_{i,j} \right) P_i + \sum_{\substack{j=1 \\ i \neq j}}^{21} \left( \gamma_{i,j}^{1+} + \gamma_{i,j}^{1-} + \gamma_{i,j}^{2+} + \gamma_{i,j}^{2-} + w_{i,j} \right) P_j, \quad (3)$$

where $P_i$ is the population of the energy eigenstate $|i\rangle$ with energy $\varepsilon_i$. The spin-phonon transition rates were calculated using a golden-rule method following Leuenberger and Loss [13]:

$$\gamma_{i,j}^{1\mp} = \frac{g_o^2}{48\pi\rho c_s^5 \hbar^4} |\langle i|\{S_\mp, S_z\}|j\rangle|^2 \frac{(\varepsilon_i - \varepsilon_j)^3}{e^{\frac{\varepsilon_i - \varepsilon_j}{T}} - 1} \quad (4)$$

$$\gamma_{i,j}^{2\mp} = \frac{g_o^2}{32\pi\rho c_s^5 \hbar^4} \left|\langle i|S_\mp^2|j\rangle\right|^2 \frac{(\varepsilon_i - \varepsilon_j)^3}{e^{\frac{\varepsilon_i - \varepsilon_j}{T}} - 1} \quad , \tag{5}$$

where $c_s$ is the sound velocity and $\rho$ the mass density. The spin-phonon coupling constant $g_o$ was determined empirically by fitting AC susceptibility data (not shown).

The radiation-induced transition rates are similarly calculated using a standard expression from electron spin resonance [14]:

$$w_{i,j} = \frac{(H_1 g \mu_B)^2}{2\hbar^2} \left|\langle i|\cos\alpha S_x + i\sin\alpha S_y|j\rangle\right|^2 \frac{T_2}{1 + \left(2\pi\nu - \frac{(\varepsilon_i - \varepsilon_j)}{\hbar}\right)^2 T_2^2} , \tag{6}$$

where $H_1$ is the magnitude of the radiative magnetic field and $T_2$ is the spin's transverse relaxation time. The ellipticity of the radiation is defined as $\tan\alpha$.

To find the steady-state magnetization in the presence of radiation, we numerically solve the 21 equations implicit in Eq. (3), setting the left side of each to be zero to determine each $P_i^{eq}$, the equilibrium population of level $|i\rangle$. From this we solved for the magnetization $M$ using

$$M = \sum_{i=1}^{21} \langle i|\vec{S} \cdot \frac{\vec{H}}{|\vec{H}|}|i\rangle P_i^{eq} , \qquad (7)$$

where $\vec{S} \cdot \frac{\vec{H}}{|\vec{H}|}$ is the spin operator along the external field direction.

In our calculations we fixed the anisotropy parameters to currently accepted values. The only parameters we varied were $H_1$, the magnitude of the radiation field, $T_2$, the spin's transverse relaxation time and (only for the results in Fig. 2a) the ellipticity of the radiation. $H_1$ only sets the amplitude of the peaks, while $T_2$ determines the width. The ellipticity controls the relative height of the two peaks. Our fits determined $H_1$ to be in the range 0.03 – 0.165 Oe, depending on frequency. In our simulations, we used $T_2$ = 0.17 ns, a value that allowed us to reproduce most of the data curves well. This value is a lower bound for $T_2$ since it may reflect the effects of inhomogeneities from dipole fields, anisotropy parameters and $g$ factors [15-17]. Our line widths are consistent with those found spectroscopically by others [18].

Our results suggest that single-molecule magnets can be employed in a novel form of magnetic storage. Instead of using an applied magnetic field to flip a bit, as is done in usual forms of magnetic storage, radiation of an appropriate frequency can be used to drive the spin from one orientation to another. In addition, these results have implications for the use of molecular magnets as qubits. By using pulsed radiation,

single- and multiple-qubit operations should be achievable. While there are other magnetic systems in which radiation can change [19] or induce [20] a magnetic state, to our knowledge the single-molecule magnets are the only bistable magnetic systems in which radiation can drive a substantial magnetization change through a quantum resonant process.

We thank M. P. Sarachik, S. Hill and D. Candela for useful conversations and advice. Millitech, Inc. kindly loaned some of the equipment used for this study. M. Tuominen generously allowed us to use some of his laboratory facilities. We are indebted to D. Krause and P. Grant for their technical contributions and advice. We also thank B. Lyons, D. Orbaker, D. Vu and M. Willis for their various contributions to this project. Support for this work was provided by the US National Science Foundation, the Research Corporation, the Alfred P. Sloan Foundation, the Center of Excellence of the Israel Science Foundation, and the Amherst College Dean of Faculty through a grant from the Andrew W. Mellon Foundation.

* Present address: Department of Physics, Amherst College, Amherst, Massachusetts 01002-5000, USA.

† Correspondence and requests for materials should be addressed to J.R.F. (jrfriedman@amherst.edu).

**Figure Captions**

Figure 1. Double-well potential and energy levels for the $Fe_8$ magnet. The left well corresponds to the spin pointing "up" and the right corresponds to it pointing "down". The photon-induced magnetization reversal process is illustrated schematically by the arrows. Resonant microwave radiation drives some molecules from the ground state to the first excited state in the right well (wavy arrow). Some of this increased population is distributed to the left well by thermal activation (red arrows), which involves multiple phonon transitions, tunneling (green arrows), which is only significant when levels in opposite wells align, or some combination of both.

Figure 2. Photon-induced magnetization change as a function of magnetic field. The induced magnetization change $\Delta M$ is normalized by the saturation magnetization $M_{sat}$. Microwave frequencies used were a) 120 GHz, b) 117.9 GHz, c) 115.5 GHz and d) 114 GHz. Peaks/dips occur when the energy between the two lowest levels in the right well (Fig. 1) matches the photon energy. The solid curves are the results of simulations, as discussed in the text. The asymmetry in peak heights in a) can be accounted for by assuming that the radiation is elliptically polarized with ellipticity 0.16.

Figure 3. Peak positions versus applied radiation frequency. The fields at which photon-induced magnetization reversal occurs are extracted from the data shown in

**Fig.2 (as well as others not shown) and plotted as a function of microwave frequency. The linear dependence derives from the Zeeman term, Eq. 2. The solid line is calculated from using the accepted Hamiltonian for the system and setting $\theta$ = 34º.**

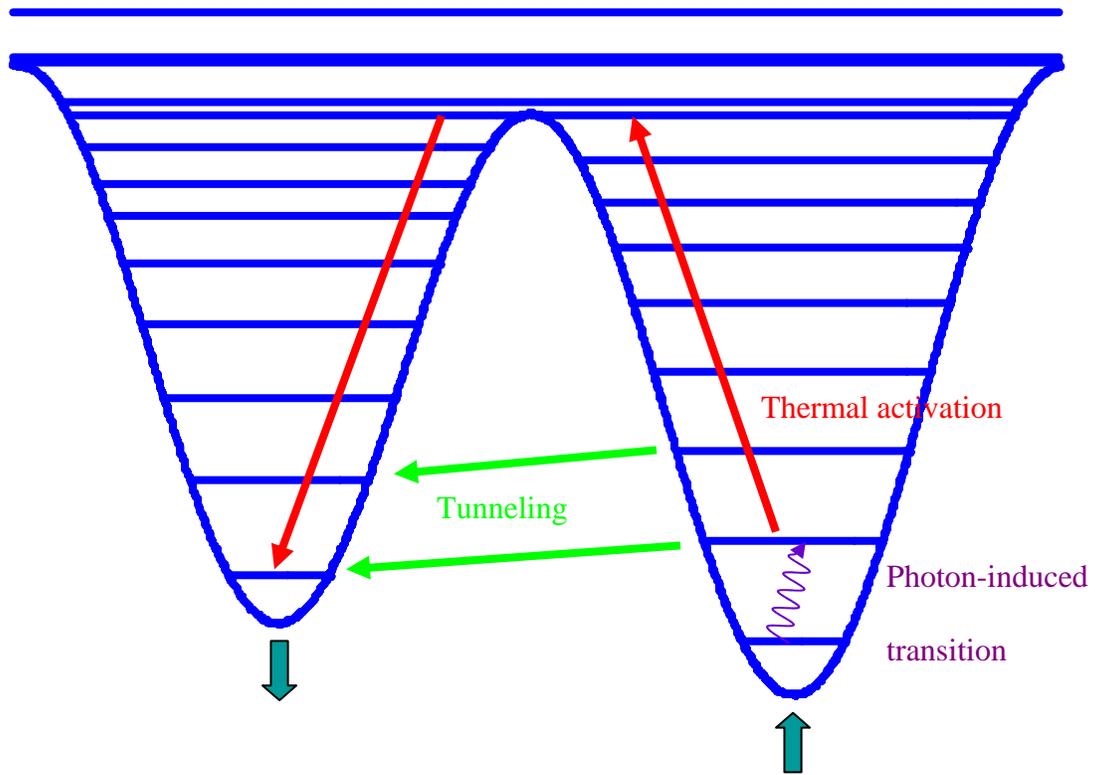

**Figure 1**

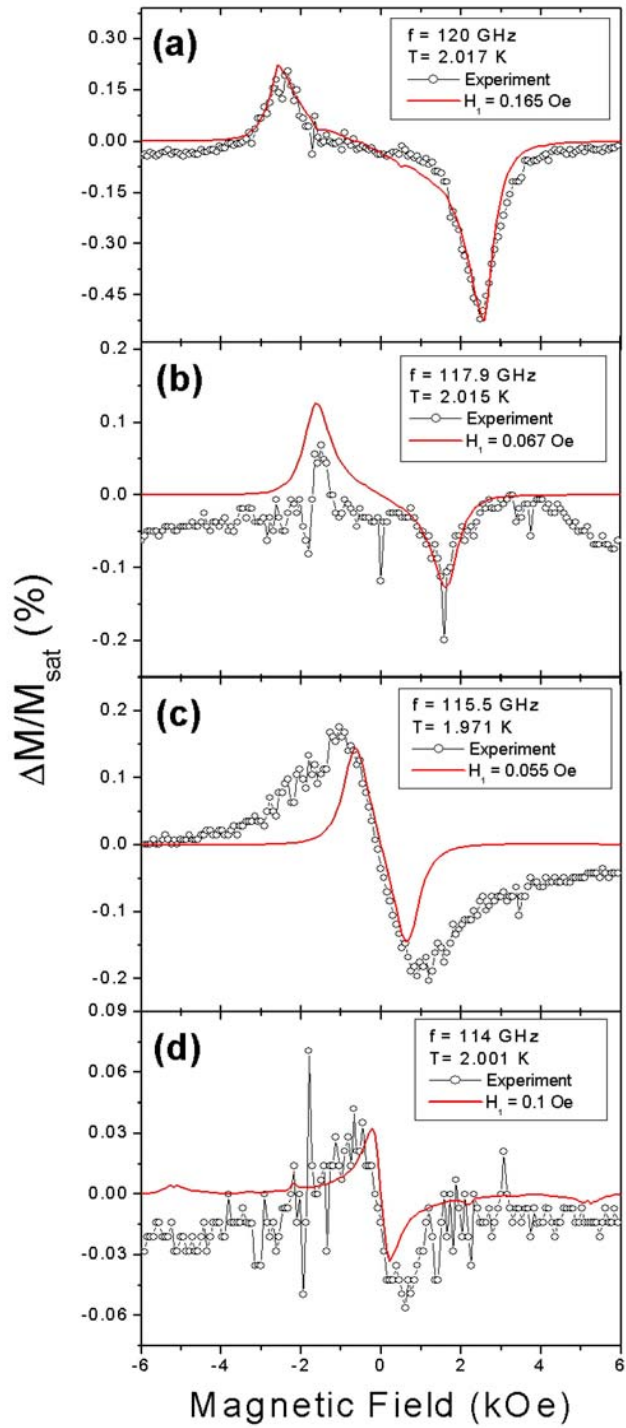

**Figure 2**

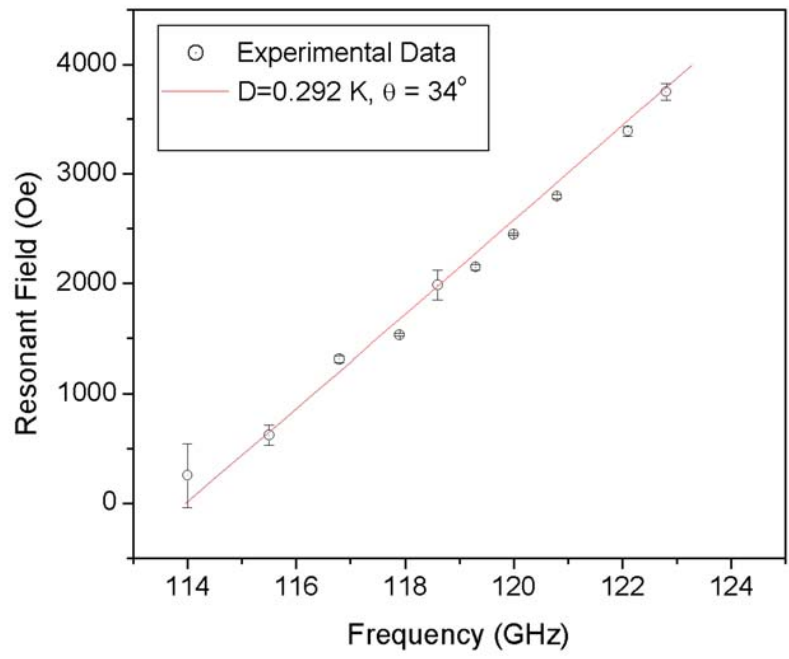

**Figure 3**